\documentclass[12pt]{article}

\usepackage{epsfig,layout}
\begin{document}

\title{
Compression-based methods  for nonparametric   density estimation,
on-line prediction, regression and classification for time series.
}

\author{Boris Ryabko
\thanks{Research  was supported  by  Russian Foundation for Basic Research (grant no. 06-07-89025-a.).}
}

\date{}
\maketitle

\begin{abstract}

We address the  problem of nonparametric estimation  of characteristics for stationary
and ergodic  time series.
We consider   finite-alphabet time series and   real-valued ones
and  the  following four problems:
  i) estimation of the (limiting) probability $P(u_0 \ldots  u_s) $ for every $s$ and each sequence $ u_0 \cdots
u_s$ of letters from the process alphabet  (or estimation of the
density $p(x_0, \ldots, x_{s}) $  for real-valued time series),
ii) so-called on-line prediction, where the conditional
probability
 $P(x_{t+1}/ x_1x_2 \ldots x_t)$  (or the conditional density
$p(x_{t+1}/ x_1x_2 \ldots x_t)$) should be estimated  (in the case
where $x_1x_2 \cdots x_t$ is known), iii) regression and iv)
classification (or so-called problems with side information).

We show that  so-called archivers (or data compressors)  can be
used as a tool for solving these problems.   In particular,
firstly, it is proven that any so-called universal code (or
universal data compressor) can be used as a basis for constructing
asymptotically optimal methods for the above  problems. (By
definition,
 a universal code can "compress"
any sequence generated by a stationary and ergodic source
asymptotically till the Shannon entropy of the source.) And, secondly,
we show experimentally  that estimates, which are based on
practically used methods of data compression,   have a reasonable
precision.

\end{abstract}

\textbf{keywords:} time series, nonparametric estimation, prediction, universal coding, data compression,
on-line prediction, Shannon entropy, stationary and ergodic process,
regression.


\section{Introduction}

We consider a stationary and ergodic source,  which generates
sequences $x_1x_2\cdots$ of elements (letters) from some set
(alphabet) $A$, which is either finite or real-valued. It is
supposed that the  probability distribution (or distribution of
limiting probabilities) $P(x_1 = a_{i_1}, x_2= a_{i_2}, \ldots,
x_t= a_{i_t}) $ (or the density $p(x_1, x_2, \ldots, x_t)$) is
unknown, but we are given  either  one sample $x_1 \ldots x_t$
 or  several ($r$) non-overlapping  samples $x^1=
x^1_1 \ldots x^1_{t_1}, $ $\ldots,$ $x^r= x^r_1 \ldots x^r_{t_r} $
generated by the source. (Here non-overlapping means that the
sequences either  are parts of deferent realizations or belongs to
non-overlapping  parts of one realization, say, a realization with
gaps.  Generally speaking, they cannot be combined into one sample
for a stationary and ergodic source, as it can be done for an
i.i.d. one.)

Of course,  if someone knows the probability distribution (or the density)
he has all information
about the source
and can solve all problems in the best way. Hence,  generally
speaking,  precise estimations of the probability distribution and
the density can be used for prediction, regression estimation,
etc. In this paper we  follow the scheme:
  we consider the problems of estimation of the probability distribution
  or
the density estimation. Then we show how the solution  can be
applied to other problems, paying the main attention to the
problem of prediction,  because of its practical applications and
importance for probability theory, information theory, statistics
and other theoretical sciences, see \cite{Al, GMY, JS,  Ki, MM,
Mo,  Ri, Sp}.

We show that  universal codes (or data compressors) can be applied
directly to the problems of estimation, prediction, regression and
classification.   It is not surprising, because for any stationary
and ergodic source $p$ generating letters from a finite alphabet
and any universal code $U$ the following equality is valid with
probability 1:
$$
        \lim_{t\rightarrow\infty}
\frac {1}{t} ( -   \log p(x_1 \cdots x_t) -  | U(x_1 \cdots x_t)|)   \:
=  \, 0 ,
$$
 where  $ x_1 \cdots x_t$ is generated by $p$.
(Here and below $\log = \log_2$,  $|v|$ is the length of $v$, if
$v$ is a word and the number of elements of $v$ if $v$ is a set.)
So, in fact,   the length of the universal code ($ | U(x_1 \cdots
x_t)|$) can be used as an estimate of the logarithm of the unknown
probability and, obviously, $2^{- | U(x_1 \cdots x_t)|} $ can be
considered as the estimation of    $ p(x_1 \cdots x_t) .$ In fact,
a universal code can be viewed as a non-parametrical estimation of
(limiting) probabilities for stationary and ergodic sources. This
was recognized shortly after the discovery of universal codes (for
the set of stationary and ergodic processes with finite alphabets
\cite{Ry0}) and universal codes were applied for solving
 prediction problem  \cite{Ry1}.

 We would like to emphasize that, on the one hand, all
results are obtained
 in the framework of
classical probability theory and mathematical statistics and, on
the other hand, everyday methods of data compression (or
archivers) can be used as a tool for density estimation,
prediction and other problems, because they are practical
realizations of universal codes.  It is worth noting  that the
modern data compressors (like \emph{zip}, \emph{arj, rar,} etc.)
are based on deep theoretical results of the theory of source
coding (see, for ex., \cite{e, K-Y,Kr,Ri,Sa}) and have been
demonstrated  high efficiency in practice as compressors of texts,
DNA sequences and many other types of real data. In fact,
 archivers can find many kinds of latent regularities, that is
why they  look like a promising tool for prediction and other
problems. Moreover, recently universal codes and archivers were
efficiently applied to some problems which are very far from data
compression: first,  their   applications in \cite{V1,V2} created
a new and rapidly growing line of investigation in clustering and
classification and, second, universal codes were used as a basis
for non-parametric tests for the main statistical hypotheses
concerned with stationary and ergodic time series \cite{RA, RM}.

The outline of the paper is as follows. The section 2 contains
description of the Laplace predictor and its generalizations, a
review of known results and  description of one universal code.
The sections 3 and 4 are devoted to processes with  finite and
real-valued alphabets, correspondingly. The last part contains
some examples and simulations.

 \section{Predictors and universal data compressors }

\subsection{The Laplace measure
and on-line prediction for i.i.d. processes  }

We consider a source with unknown statistics which generates sequences
$x_1x_2\cdots$ of letters from some set (or alphabet) $A$.
Let the source generate a message $x_1\ldots x_{t-1}x_t\ldots $,
$\, x_i\in A$ for all $i$, and the following letter $ x_{t+1}$ needs to
be predicted.

It will be convenient at first to  describe briefly the prediction
problem. This problem can be traced back to Laplace \cite{FE}. He
considered the problem how  to estimate the probability that the
sun will rise tomorrow, given that it has risen every day since
Creation. In our notation the alphabet $A$ contains two letters $
0 \; ("the \:sun \:rises") $  and $1 \;("the \:sun\: does\: not\:
rise" ) ,$ $t$ is the number of days since Creation, $x_1\ldots
x_{t-1}x_t = 00 \ldots 0 .$

Laplace suggested the following predictor:
\begin{equation}\label{L}
L_0(a|x_1\cdots x_t) = (\nu_{x_1\cdots x_t}(a) +1 )/ (t+ |A | ),
\end{equation}  see \cite{FE}, where $\nu_{x_1\cdots x_t}(a)$
denote the count of letter $a$ occurring in the word $x_1\ldots
x_{t-1}x_t.$
For example, if
$ A= \{0, 1 \}, \: x_1 ... x_5 = 01010,$ then the Laplace prediction
is as follows: $L_0(x_{6}=0| 01010) = (3+1)/ (5+2) = 4/7,
L_0(x_{6}=1 | 01010) = (2+1)/ (5+2) = 3/7.$ In other words, $3/7$
and  $4/7 $  are estimations of the unknown probabilities $P(x_{t+1}
= 0|x_1 \ldots x_t = 01010 )$ and $P(x_{t+1} = 1 |x_1 \ldots x_t =
01010 ) .$

We can see that Laplace  considered  prediction as a set of
estimations of unknown (conditional) probabilities. This approach
to the problem of prediction was developed in \cite{Ry1} and now
is
 often called on-line prediction or universal prediction \cite{Al, GMY,
 Mo}. As we mentioned above, it seems  natural to consider conditional probabilities to be
 the  best prediction, because they contain all information about
 the
 future  behavior of the stochastic process. Moreover, this approach
 is deeply connected with game-theoretical interpretation of
 prediction (see \cite{Ke, R94}) and, in fact,
 all obtained results can be easily transferred from one model to the other.

               Any predictor $\gamma$  defines a measure by following equation
\begin{equation}\label{p} \gamma (x_1 ... x_t) = \prod_{i=1}^t \gamma(x_i | x_1 ... x_{i-1}
).
\end{equation}
For example, $L_0(0101) = \frac{1}{2} \frac{1}{3} \frac{1}{2}
\frac{2}{5} = \frac{1}{30}.$
And, vice versa, any measure $ \gamma$ (or estimation of the measure)
defines a predictor: $\gamma(x_i | x_1 ... x_{i-1}) =
 \gamma( x_1 ... x_{i-1} x_i) /  \gamma( x_1 ... x_{i-1} ) .$
The same is true for a density  (and its estimation): a predictor
is defined  by conditional density and, vice versa, the density is
equal to the product of conditional densities:
$$  p(x_i | x_1 ... x_{i-1}) =  p( x_1 ... x_{i-1} x_i) /
p( x_1 ... x_{i-1} ),  p (x_1 ... x_t) =
 \prod_{i=1}^t  p(x_i | x_1 ... x_{i-1}). $$

 The next natural question is how to estimate the precision or of
 the
 prediction and an estimation of probability.
Mainly we will  estimate the error of prediction  by the
Kullback-Leibler (KL) divergence
 between a distribution $p$ and its estimation. Consider
an (unknown)  source $p$ and some predictor $\gamma$. The {\it{
error }} is characterized by the KL divergence
\begin{equation}\label{r0}
\rho_{\gamma,p}(x_1\cdots x_t)= \sum_{a\in A}p(a|x_1\cdots
x_t)\log\frac{p(a|x_1\cdots x_t)} {\gamma(a|x_1\cdots x_t)}.
\end{equation}
It is well-known that for any distributions $p$ and $\gamma$ the
K-L divergence is nonnegative and equals 0 if and only if $p(a) =
\gamma(a)$ for all $a,$ see, for ex., \cite{Ga}. The following
inequality (Pinsker's inequality)
\begin{equation}\label{pi}
\sum_{a\in A} P(a)\log\frac{P(a)} {Q(a)} \geq  \frac{\log e} {2} ||
P - Q ||^2.
\end{equation}
connects the KL divergence with a so-called variation distance $$
|| P - Q || = \sum_{a \in A} | P(a) - Q(a)| , $$ where $P$ and $Q$
are distributions over $A,$ see \cite{CK}. For fixed $t$,
$\rho_{\gamma,p}( \,)$ is a random variable, because $x_1, x_2,
\cdots, x_t$ are random variables. We define the average error
  at time $t$ by
\begin{equation}\label{r1}
\rho^t(p\|\gamma)=E\,\left(\rho_{\gamma,p}(\cdot)\right)=\nonumber
\, \sum_{x_1\cdots x_t\in A^t}p(x_1\cdots
x_t)\,\,\rho_{\gamma,p}(x_1\cdots x_t).
\end{equation}
It is shown in \cite{Ry2} that the error of Laplace predictor  $L_0$  goes
to 0 for any i.i.d. source $p$. More precisely, it is proven that
\begin{equation}\label{rL} \rho^t(p\|L_0)
< (|A| - 1)/ (t + 1) \end{equation}
 for any source $p \, ,$
 (\cite{Ry2};  see also \cite{RT}). So, we can see from this inequality that the
 average error of the Laplace predictor $L_0$ (estimated either
 by the KL divergence or the variation distance ) goes to zero for any unknown i.i.d. source,
 when the sample size $t$ grows. Moreover, it can be easily shown
 that the error (\ref{r0}) (and the corresponding variation
 distance) goes to zero with probability 1, when $t $ goes to infinity.
 Obviously, such a property is very
 desirable for any predictor and for larger classes of sources, like Markov,
  stationary and ergodic, etc.  However, it is proven in
  \cite{Ry1} (see also \cite{Al, GMY, Mo}) that such predictors do not exist for the class of all stationary and ergodic sources
  (generated letters from a given finite alphabet).
  More precisely,  for any predictor $\gamma$ there exists a source $p$ and $\delta > 0$ such that
  with probability 1 $\:\rho_{\gamma,p}(x_1\cdots
  x_t) \geq \, \delta$ infinitely often when
  $t\rightarrow \infty.$
  So,  the error of any predictor does
not go to 0, if the predictor is applied to all stationary and
ergodic sources, that is why it is difficult to use (\ref{r0}) and
(\ref{r1}) for comparison of different predictors.

On the other hand, it is shown in \cite{Ry1} that there exists
 a predictor $R$, such that the following Cesaro average $
  t^{-1}\: \sum_{i=1}^t
\rho_{R,p} (x_1\cdots x_t) $
 goes to 0 (with probability 1 ) for any stationary and ergodic source
$p,$ where  $t$ goes to infinity.  That is why we will focus our
attention on such averages and by analogy with (\ref{r1}) we define
\begin{equation}\label{R0} \bar{\rho}_{\gamma, p}(x_1 ... x_t) =
t^{-1}\: ( \log ( p (x_1 ... x_t) / \gamma (x_1 ... x_t) )
\end{equation} and
\begin{equation}\label{RR} \bar{\rho}_t (\gamma, p) = t^{-1} \sum_{x_1 ... x_t \in A^t} p(x_1 ... x_t)
\log ( p (x_1 ... x_t) / \gamma (x_1 ... x_t) ),
\end{equation} where, as before,
$
 \gamma (x_1 ... x_t) = \prod_{i=1}^t \gamma(x_i | x_1 ... x_{i-1}
).
$

From these definitions and (\ref{rL}) we obtain the following
estimation of the error of the Laplace predictor $L_0$ for any
i.i.d. source:
\begin{equation}\label{Lr}
\bar{\rho}_t (L_0, p) < ((|A| - 1)\, \log t  + c)/t,
\end{equation} where $c$ is a certain constant.
So, we can see that the average  error of the Laplace predictor
goes to zero for any i.i.d. source (which generates letters from a
known finite alphabet). As a matter of fact, the Laplace
probability $L_0( x_1 ... x_t) $ is a consistent estimate of the
unknown probability $p( x_1 ... x_t) .$

The natural problem is to find  a predictor whose error is minimal
(for  i.i.d. sources). This problem was considered and solved by
Krichevsky \cite{Kr1}, see also  \cite{Kr}. He suggested the
following predictor:
\begin{equation}\label{kr1}
K_0(a|x_1\cdots x_t) = (\nu_{x_1\cdots x_t}(a) +1/2 )/ (t+ |A |/2 ),
\end{equation}  where, as before,  $\nu_{x_1\cdots x_t}(a)$
denote the count of letter $a$ occurring in the word $x_1\ldots
 x_t.$ We can see that the Krychevsky predictor is quite
close to the Laplace's  one (\ref{L}). For example, if $ A= \{0, 1
\}, \: x_1 ... x_5 = 01010,$ then $K_0(x_{6}=0| 01010) = (3+1/2)/
(5+1) = 7/12, K_0(x_{6}=1 | 01010) = (2+1/2)/ (5+1) = 5/12$ and
$K_0(01010) = \frac{1}{2} \frac{1}{4} \frac{1}{2} \frac{3}{8}
\frac{1}{2}
 = \frac{3}{256}.$

The Krichevsky measure  $K_0$ can be presented as follows:
\begin{equation}\label{Kp}
K_0(x_1 ... x_t) =\prod_{i=1}^{t} \frac{ \nu_{x_1 ... x_{i-1}}
(x_i)+ 1/2 }{i-1+|A|/2} = \frac{\prod_{a \in A} (\prod_{j=
1}^{\nu_{x_1 ... x_t}(a)} (j- 1/2))}{\prod_{i= 0}^{t-1}(i+ |A|/2)}
\,.
\end{equation}
It is known that
\begin{equation}\label{gamma}
 (r+1/2) ((r + 1) + 1/2) ... (s - 1/2) =
\frac{\Gamma(s+1/2)}{\Gamma(r+1/2)},
\end{equation}
where
$\Gamma(\:)$ is the gamma function (see for definition, for ex.,
\cite{Kn} ). So, (\ref{Kp}) can be presented as follows:
 \begin{equation}\label{Kp1}
K_0(x_1 ... x_t) = \:
\frac{\prod_{a \in A} ( \Gamma(\nu_{x_1 ... x_t}(a)+ 1/2
) \,  / \Gamma(1/2) \,)   }{\Gamma(t+ |A|/2 ) \, / \Gamma(|A|/2)  } \,.
\end{equation}
For this predictor
\begin{equation}\label{kr2}
\bar{\rho}_t (K_0, p) < ((|A| - 1)\,  \log t  + c) /(2t),
\end{equation} where $c$ is a constant,  and, moreover,
in a certain sense this average error is   minimal:
for any predictor $\gamma$ there exists such  a source $p^*$ that
$$  \bar{\rho}_t (\gamma, p^*)  \geq ((|A| - 1)\,  \log t  + c) /(2t),
$$ see
\cite{Kr1},  \cite{Kr}.

 \subsection{ Consistent estimations and on-line predictors  for Markov and ergodic processes}

Now we briefly describe consistent estimations of unknown probabilities and efficient on-line predictors  for general  stochastic processes (or sources of
information).  Denote by $A^t$ and $A^*$ the set of
all words of length $t$ over $A$ and the set of all finite
words over $A$ correspondingly ($A^* = \bigcup_{i=1}^\infty
A^i$). By
$M_\infty(A)$ we denote the set of all stationary and ergodic
sources, which generate letters from $A$ and let $M_0(A) \subset M_\infty(A)$ be
the set of all i.i.d. processes.
 Let $M_m(A) \subset M_\infty(A)$ be the set of Markov
sources of order (or with memory, or connectivity) not larger than
$m, \, m \geq 0. $   Let
$M^*(A) = \bigcup_{i=0}^\infty M_i(A)$ be the set of all
finite-order sources.

The Laplace and Krichevsky predictors can be extended to
general Markov processes.
  The trick
is to view a Markov source $p\in M_m(A)$ as resulting from $|A|^m$
i.i.d. sources. We illustrate this idea by an example from
\cite{RT}. So assume that $A=\{O,I\}$, $m=2$ and assume that the
source $p\in M_2(A)$ has generated the sequence
$$ OOIOIIOOIIIOIO. $$
We represent this sequence by the following four subsequences:
$$ **I*****I***** ,$$
$$ ***O*I***I***O ,$$
$$ ****I**O****I* ,$$
$$ ******O***IO** .$$
These four subsequences contain letters which follow
$OO$,  $OI$,  $IO$ and  $II$, respectively.
 By definition,
$p\in M_m(A)$ if $p(a |x_1\cdots x_t) = p(a|x_{t-m+1}\cdots x_t
$), for all $0 < m \leq t $, all $a\in A$ and all $x_1\cdots x_t
 \in A^t $. Therefore, each of the four generated subsequences
may be considered to be generated by a Bernoulli source. Further,
it is possible to reconstruct the original sequence if we know the
four ($=|A|^m$) subsequences and the two ($=m$) first letters of
the original sequence.

Any predictor $\gamma $ for i.i.d. sources can be applied for
Markov sources. Indeed, in order to predict, it is enough to store
in the memory $|A|^m$ sequences, one corresponding to each word in
$A^m$. Thus, in the example, the letter $x_3$ which follows
$OO$ is predicted based on the Bernoulli method $\gamma$
corresponding to the $x_1 x_2$- subsequence ($=OO$),
 then $x_4$ is predicted based on the Bernoulli
method corresponding to $x_2x_3$, i.e. to the $OI$-
subsequence, and so forth. When this scheme is applied along
with either  $L_0$ or  $K_0$ we denote the obtained predictors
as
$L_m$ and $K_m,$ correspondingly and define the probabilities
for the first $m$ letters as follows: $ L_m(x_1) = L_m(x_2) =
\ldots = L_m (x_m) = 1/|A|\,, $ $K_m(x_1) = K_m(x_2) = \ldots =
K_m(x_m) = 1/|A|\,. $   For example, having taken into account
 (\ref{Kp1}), we can present the Krichevsky predictors for $M_m(A)$ as follows:
\begin{equation}\label{km}
K_m(x_1 ... x_t) =\cases{\frac{1}{|A|^t},&if $t \leq m\, ,$\cr
 & \cr
\frac {1} {|A|^{m} }  \,  \prod_{v \in A^m} \frac{\prod_{a \in A}\:
           (( \Gamma( \nu_x(v a )+ 1/2)  \, / \, \Gamma(1/2)) }
{( \Gamma( \bar{\nu}_x(v  )+|A|/2) \,  / \,  \Gamma(|A|/2) )}, &if $ t > m $ \, ,}
\end{equation}
where $\bar{\nu}_x(v  )= \sum_{a \in A} \nu_x(v a ), \:x = x_1 ...
x_t. $
It is worth noting that the
representation (\ref{gamma})
can be more convenient for carrying out
calculations.  Let us consider an example. For the word
$ OOIOIIOOIIIOIO $ considered in the previous example,    we obtain
$
K_2(OOIOIIOOIIIOIO) = $ $   2^{-2}   \,\, \,   \frac{1} {2} \frac{3} {4}
 \,\,  \frac{1} {2} \frac{1} {4} \frac{1} {2} \frac{3} {8}  \, \, \,
 \frac{1} {2} \frac{1} {4} \frac{1} {2} \, \, \,
 \frac{1} {2} \frac{1} {4} \frac{1} {2} \, \, .
$

Let us define the measure $R,$  which, in fact,  is a consistent
estimator of probabilities for the class of all stationary and
ergodic processes with a finite alphabet. First we define
 a probability distribution $\{\omega =
\omega_1, \omega_2, ... \}$ on integers $\{ 1, 2, ... \}$ by
\begin{equation}\label{om} \omega_1 = 1 - 1/ \log 3,\: ... \,,\:
\omega_i\,= 1/ \log (i+1) - 1/ \log (i+ 2),\: ... \; .
\end{equation}
(In what follows we will use this distribution, but
results  described below are obviously true for   any
distribution with nonzero probabilities.)
The measure $R$
is defined as follows:
\begin{equation}\label{R}
 R(x_1 ... x_t) = \sum_{i=0}^\infty \, \omega_{i+1} \:K_i(x_1 ...
 x_t) . \end{equation}
 It is worth
 noting that this construction can be applied to the Laplace
 measure
 (if we use $L_i$ instead of $K_i$) and any other family of measures.

 The main properties of the measure $R$  are connected with the
Shannon entropy, which is defined as follows
\begin{equation}\label{H}
 H(p) =  \lim_{m \rightarrow \infty}  -    \frac {1}{m}
\sum_{v \in A^m} p(v) \log p(v).
\end{equation}


\textbf{Theorem 1. }  \cite{Ry1}.  \emph{For any stationary and ergodic source $p$  the following
equalities are valid:}
$$  i) \, \,    \lim_{t\rightarrow\infty}
\frac {1}{t}    \log (1/ R(x_1 \cdots x_t) )     \: =  \, H(p)
$$ \emph{  with probability 1, }
$$ ii) \, \, \, \,   \lim_{t\rightarrow\infty}
\frac {1}{t}  \sum_{u \in A^t} p(u)  \log ( 1 / R (u ))   \:
=  \, H(p) .
$$


\subsection{  Nonparametric estimations and data compression}

One of the goals of the paper is to show how practically used
data compressors can be used as a tool for nonparametric estimation,
prediction and other problems. That is why a short description
of universal data compressors (or universal codes)  will be given here.

 A data compression method (or code) $\varphi$ is defined as a
set of mappings $\varphi_n $ such that $\varphi_n : A^n
\rightarrow \{ 0,1 \}^*,\, n= 1,2, \ldots\, $ and for each pair of
different words $x,y \in A^n \:$ $\varphi_n(x) \neq \varphi_n(y)
.$  It is also required that each sequence
$\varphi_n(u_1)\varphi_n(u_2) ...\varphi_n(u_r), r \geq 1,$ of
encoded words from the set $A^n, n\geq 1,$ could be uniquely
decoded into $u_1u_2 ...u_r$. Such codes are called uniquely
decodable. For example, let $A=\{a,b\}$, the code $\psi_1(a) = 0,
\psi_1(b) = 00, $  obviously, is not uniquely decodable. It is
well known
 that if  a code $\varphi$ is uniquely decodable
then  the lengths of the codewords satisfy the following
inequality (Kraft's inequality): $ \sum_{u \in A^n}\: 2^{-
|\varphi_n (u) |} \leq 1\:,$ see, for ex., \cite{Ga}. It will be
convenient to reformulate this property as follows:

\textbf{Claim 1.} \emph{Let $\varphi$ be a uniquely decodable code over
an alphabet $A$. Then  for any integer $n$ there exists a measure
$\mu_\varphi$ on $A^n$ such that
\begin{equation}\label{kra}
 - \log \mu_\varphi (u) \:  \leq  \: |\varphi (u)|
\end{equation} for any $u$ from $A^n \,.$}

(Obviously, Claim 1 is true for the measure $
 \mu_\varphi (u) =
2^{- |\varphi (u) |} / \Sigma_{u \in A^n}\: 2^{- |\varphi (u)|}
).$
 In what follows we call
 uniquely decodable codes just "codes".

It is worth noting that, in fact, any measure $\mu$   defines
a code for which the length of
 the codeword associated with a word $u$ is  (close to)  $- \log \mu(u).$

Now we consider universal codes.  By definition, a code $U$ is universal
if for any stationary and ergodic source $ p $ the following equalities
are valid:

\begin{equation}\label{U1}
\lim_{t\rightarrow\infty}  |U(x_1 \ldots x_t) |  /t =
H(p)
\end{equation}  with probability 1, and
\begin{equation}\label{U2}
\lim_{t\rightarrow\infty}   E ( |U(x_1 \ldots x_t) |  ) /t =
H(p) ,
\end{equation}   where $H(p)$ is the Shannon entropy of $p,$
$E( f)$  is a mean value of  $f$. In fact, (\ref{U2}) and
(\ref{U1}) are valid for known  universal codes, but there exist
codes for which only one equality is valid.

\section{Finite-alphabet processes}
\subsection{ The estimation  of (limiting) probabilities }

The following theorem shows
how universal codes can be applied for probability estimations.

\textbf{Theorem 2. }  \emph{Let $U$ be a universal code and }

\begin{equation}\label{UC}    \mu_U (u) =
2^{- |U (u) |} / \Sigma_{ v \in A^{|u|}}\: 2^{- |U(v)|}
.
\end{equation}  \emph{
Then, for any stationary and ergodic source $p$  the following equalities
are valid:  }
$$  i) \, \,    \lim_{t\rightarrow\infty}
\frac {1}{t} ( -   \log  p(x_1 \cdots x_t) -  (-  \log  \mu_U(x_1 \cdots x_t)  ) ) \:
=  \, 0
$$ \emph{  with probability 1, }
$$ ii) \, \, \, \,   \lim_{t\rightarrow\infty}
\frac {1}{t}  \sum_{u \in A^t} p(u)  \log ( p(u) / \mu_U (u ))   \:
=  \, 0 ,
$$
$$  iii)  \, \, \, \,   \lim_{t\rightarrow\infty}
  \frac {1}{t}
 \sum_{u \in A^t} p(u) \,  |p(u)  - \mu_U (u )|  \:
=  \, 0.  $$

\emph{ Proof} is based on
 Shannon-MacMillan-Breiman Theorem which states that
 for any stationary and ergodic source $p$ $$
\lim_{t\rightarrow\infty} -  \log p(x_1 \ldots x_t) /t =
H(p)
$$
 with probability 1, see \cite{Billingsley, Ga}.
From this equality and  (\ref{U1}) we obtain the statement i). The
second statement follows from the definition of Shannon entropy
(\ref{H}) and (\ref{U2}), whereas iii) follows from ii) and the
Pinsker's  inequality (\ref{pi}).

So, we can see that, in a certain sense,
the measure $\mu_U$ is a consistent (nonparametric) estimation of
the (unknown)  measure $p.$

Nowadays there are many efficient universal codes (and universal
 predictors connected with them), see \cite{JS,Ki,
 No,Ri,Ry1,Sa}, which can be applied to estimation.
For example, the above described measure $R$ is based on the code
from  \cite{Ry0,Ry1} and can be applied for probability
estimation. More precisely, Theorem 2 (and the following theorems)
are true for $R$, if we replace $\mu_U$ by $R.$

 It is important to note that the  measure $R$ has some additional
properties, which can be useful for  applications. The following
theorem will be devoted to description  of these properties
(whereas all other theorems are valid for all universal codes and
corresponding them measures, including the measure $R ).$

\textbf{Theorem 3.}  \emph{For any Markov process $p$ with memory $k$}

\emph{i) the error of the probability estimator, which is based on the measure $R,$ is upper-bounded as follows:}
$$   \frac {1}{t}  \sum_{u \in A^t} p(u)  \log ( p(u) / R(u ))   \: \leq   \frac {(|A|-1) |A|^{k-1}  \log t} {2\,t} + O(\frac {1}{t}  ),
$$

\emph{ii) in a certain sense the error of $R$ is asymptotically minimal: for any measure $\mu$
there exists a $k-$memory Markov process $p_\mu$ such that}
$$   \frac {1}{t}  \sum_{u \in A^t} p_\mu(u)  \log ( p_\mu(u) / \mu(u ))   \: \geq  \frac {(|A|-1) |A|^{k-1}  \log t} {2\,t} + O(\frac {1}{t}  ),
$$

iii) \emph{Let $\Theta$ be such a set of stationary and ergodic processes  that there  exists a measure $\mu_\Theta$ for which the    estimation error of the probability goes to 0 uniformly:}
$$  \lim_{t\rightarrow\infty} \; \sup_{p \in \Theta} \: (\, \frac {1}{t}  \sum_{u \in A^t} p(u)  \log ( p(u) / \mu_\Theta(u ))\, )= 0.$$ \emph{Then the error of estimator, which is based on the measure $R,$ goes to 0 uniformly, too:}
$$  \lim_{t\rightarrow\infty} \; \sup_{p \in \Theta} \: ( \,\frac {1}{t}  \sum_{u \in A^t} p(u)  \log ( p(u) / R(u ))\, )= 0.$$
\emph{Proof} can be found in \cite{Ry1,Ry2}.

\subsection{ Prediction}

As we mentioned above, any universal code $U$ can be applied for prediction.
Namely, the  measure $\mu_U$   (\ref{UC}) can be used for prediction
as the following conditional probability:
\begin{equation}\label{preu}
\mu_U(x_{t+1}| x_1 ... x_t) =  \mu_U(x_1 ... x_t x_{t+1})/ \mu_U( x_1 ... x_t).
\end{equation}

\textbf{Theorem 4. } \emph{Let $U$ be a universal code and $p$ be any stationary
and ergodic process. Then}
$$
i)  \,   \lim_{t\rightarrow\infty}   \frac {1}{t}   \, \,
\{E( \log  \frac{ p(x_1) } { \mu_U(x_1) }) +
E(\log  \frac{ p(x_2|x_1) } { \mu_U(x_2|x_1) } )+  \ldots +
E(\log  \frac{ p(x_{t}|x_1 ...
 x_{t-1}) } { \mu_U(x_{t}|x_1 ... x_{t-1}) })
\} \: = \,0, $$
$$ ii)   \,   \lim_{t\rightarrow\infty} E(  \frac {1}{t}   \, \,
\sum_{i=0}^{t-1} ( p(x_{i+1}|x_1 ... x_i)  - \mu_U(x_{i+1}|x_1 ...
x_i) )^2 )\:=\,0\,,$$\emph{ and}
$$ iii)   \,   \lim_{t\rightarrow\infty} E(  \frac {1}{t}   \, \,
\sum_{i=0}^{t-1} | p(x_{i+1}|x_1 ... x_i)  - \mu_U(x_{i+1}|x_1 ...
x_i) | )\:=\,0\,.$$

\emph{Proof}   i) immediately follows from the second statement of
the previous theorem and properties of $\log$.  The statement ii)
can be proved as follows:

$$    \,   \lim_{t\rightarrow\infty}
E(  \frac {1}{t}   \, \,   \sum_{i=0}^{t-1} ( p(x_{i+1}|x_1 \ldots  x_i)  -
\mu_U(x_{i+1}|x_1 \ldots x_i) )^2)  \leq  $$
$$\lim_{t\rightarrow\infty} E(  \frac {1}{t}   \, \,   \sum_{i=0}^{t-1} ( \sum_{a \in A}
|p(a|x_1 \ldots x_i)  - \mu_U(a|x_1 \ldots x_i)| )^2) \leq  $$
$$\lim_{t\rightarrow\infty} E(  \frac {const}{t}   \, \,   \sum_{i=0}^{t-1}
 \sum_{a \in A} p(a|x_1 \ldots x_i)  \log (p(a|x_1 \ldots x_i)  /
\mu_U(a|x_1 \ldots x_i) ) )  = $$
$$\lim_{t\rightarrow\infty} (  \frac {const}{t}   \, \,   \sum_{i=0}^{t-1} p(x_1 \ldots x_i)
 \sum_{a \in A} p(a|x_1 \ldots x_i)  \log (p(a|x_1 \ldots x_i)  /
\mu_U(a|x_1 \ldots x_i) ) )  = $$
$$ \lim_{t\rightarrow\infty}(  \frac {const}{t}   \, \,   \sum_{ x_1 \ldots x_t  \in A^t}
p(x_1 \ldots x_t)  \log (p(x_1 \ldots x_t)  / \mu(x_1 \ldots x_t)
) ). $$ Here the first inequality is obvious, the second follows
from the Pinsker's inequality (\ref{pi}), the others from
properties of expectation  and $\log .$ iii) can be derived from
ii) and the Jensen inequality for the function $x^2. $ Theorem is
proven.

\emph{Comment 1.} The  measure $R$ described above has one
additional property, if it is used for prediction. Namely, for any
Markov process $p$ ($p \in M^*(A)$ ) the following is true:
$$  \,   \lim_{t\rightarrow\infty}     \, \,\log  \frac{ p(x_{t+1}|x_1 ...
 x_t) } { R(x_{t+1}|x_1 ... x_t)}
=0 $$ with probability 1, where $R(x_{t+1}|x_1 ... x_t)=R(x_1 ... x_tx_{t+1})/R(x_1 ... x_t);$ see \cite{Ry2}.

\emph{Comment 2.} In fact, the statements ii) and iii) are
equivalent, because one of them follows from the other. For
details see Lemma~2 in \cite{DR}.

\subsection{ Problems with side information }
Now we consider  so-called problems with side information, which are described as follows: there is a stationary and ergodic source, whose
 alphabet $A$ is presented as a product $ A = X \times Y.$ We are given a sequence $  (x_1,y_1), \ldots, $ $ (x_{t-1},y_{t-1})$ and
 so-called side information $y_t.$ The goal is to predict, or estimate, $x_t.$
 This problem arises in statistical decision theory, pattern recognition, and machine learning, see \cite{Mo}.
 Obviously, if someone knows the conditional probabilities $p(x_t| $ $(x_1,y_1), \ldots, $ $ (x_{t-1},y_{t-1}), y_t)$
 for all $x_t \in X, $ he has all  information  about $x_t,$ available before $x_t$ is known.
 That is why  we will look for the best (or, at least, good) estimations for this conditional probabilities.
 Our solution will be based on results obtained in two previous subparagraphs. More precisely, for any
 universal code $U$ and the corresponding measure $\mu_U$ (\ref{UC}) we define the following estimate
 for the problem with side information:$$
\mu_U(x_t| (x_1,y_1), \ldots, (x_{t-1},y_{t-1}), y_t) =
\frac{\mu_U( (x_1,y_1), \ldots, (x_{t-1},y_{t-1}), (x_t,y_t)) }
{\sum_{x_t \in X}\mu_U( (x_1,y_1), \ldots, (x_{t-1},y_{t-1}), (x_t,y_t))}.$$

\textbf{Theorem 5. } \emph{Let $U$ be a universal code and $p$ be any stationary
and ergodic process. Then}
$$
i)  \,   \lim_{t\rightarrow\infty}   \frac {1}{t}   \, \, \{E(
\log  \frac{ p(x_1|y_1) } { \mu_U(x_1|y_1) }) + E(\log  \frac{
p(x_2|(x_1,y_1),y_2) } { \mu_U(x_2|(x_1,y_1),y_2) }) +  \ldots $$
$$+ E(\log  \frac{ p(x_{t}|(x_1,y_1), ..., (x_{t-1},y_{t-1}), y_t)
} { \mu_U(x_{t}|(x_1,y_1), ..., (x_{t-1},y_{t-1}), y_t) }) \} \: =
\,0 ,$$
$$ ii)   \,   \lim_{t\rightarrow\infty} E(  \frac {1}{t}   \, \,
\sum_{i=0}^{t-1} ( p(x_{i+1}|(x_1,y_1), ..., (x_{i},y_{i}),
y_{i+1}))  - $$ $$ \mu_U(x_{i+1}|(x_1,y_1), ..., (x_{i},y_{i}),
y_{i+1}))^2 )\: \quad=\,0\,,$$ \emph{ and}
$$ iii)   \,   \lim_{t\rightarrow\infty} E(  \frac {1}{t}   \, \,
\sum_{i=0}^{t-1} | p(x_{i+1}|(x_1,y_1), ..., (x_{i},y_{i}),
y_{i+1}))  - $$ $$ \mu_U(x_{i+1}|(x_1,y_1), ..., (x_{i},y_{i}),
y_{i+1})| )\: \quad=\,0\,.$$

\emph{Proof.} The following inequality follows from the
nonnegativity of the K-L divergency (see (\ref{pi})), whereas
equality is obvious.
$$ E( \log  \frac{ p(x_1|y_1) } { \mu_U(x_1|y_1) }) +
E(\log  \frac{ p(x_2|(x_1,y_1),y_2) } { \mu_U(x_2|(x_1,y_1),y_2) }) +  \ldots \leq $$
$$E( \log  \frac{ p(y_1) } { \mu_U(y_1) }) + E( \log  \frac{ p(x_1|y_1) } { \mu_U(x_1|y_1) })+
E(\log  \frac{ p(y_2|(x_1,y_1) } { \mu_U(y_2|(x_1,y_1) }) + E(\log  \frac{ p(x_2|(x_1,y_1),y_2) } { \mu_U(x_2|(x_1,y_1),y_2) }) +  \ldots  $$
$$ = E( \log  \frac{ p(x_1,y_1) } { \mu_U(x_1,y_1) }) + E(\log  \frac{ p((x_2,y_2)|(x_1,y_1)) } { \mu_U((x_2,y_2)|(x_1,y_1)) }) + ... .$$ Now we can apply the first statement of the theorem 4 to the last sum as follows:
$$ \lim_{t\rightarrow\infty}   \frac {1}{t} E( \log  \frac{ p(x_1,y_1) } { \mu_U(x_1,y_1) }) + E(\log  \frac{ p((x_2,y_2)|(x_1,y_1)) } { \mu_U((x_2,y_2)|(x_1,y_1)) }) + ... $$
$$ E(\log  \frac{ p((x_t,y_t)|(x_1,y_1)\ldots (x_{t-1},y_{t-1})) } { \mu_U((x_t,y_t)|(x_1,y_1)\ldots (x_{t-1},y_{t-1})) }) = 0.$$
From this equality and last inequality we obtain the proof of i).
The proof of the second statement can be obtained from the similar
representation for ii) and the second statement of the theorem 4.
iii) can be derived from ii) and the Jensen inequality for the
function $x^2.$ Theorem is proven.

\subsection{ The case of several independent samples   }

Now we extend our consideration  to the case where the sample is
presented as several non-overlapping sequences $x^1= x_1^1 \ldots
x_{t_1}^1,$ $x^2 = x_1^2 \ldots x_{t_2}^2, ... ,$ $x^r = x_1^r
\ldots x_{t_r}^r $ generated by a source. More precisely, we will
suppose that all sequences were  created by one stationary and
ergodic source. (As it was mentioned above,  it is impossible just
to combine all samples into one, if the source  is not i.i.d.) We
denote this sample by $x^1 \diamond x^2 \diamond \ldots \diamond
x^r$ and define
 $\nu_{x^1\diamond
x^2\diamond ... \diamond x^r } (v) = \sum_{i=1}^r \nu_{x^i}(v) .$
For example, if $x^1 = 0010, x^2 = 011,$ then $\nu_{x^1\diamond
x^2}(00)= 1.$ The definition of  $K_m$ and $R$ can be extended to
this case:
\begin{equation}\label{km1}
K_m(x^1\diamond x^2\diamond ... \diamond x^r  ) =  \end{equation}
 $$ (
\prod_{i=1}^r |A|^{\, -  \min{ \{m, t_i\}    } }  \,  ) \, \, \,
\, \prod_{v \in A^m} \frac{\prod_{a \in A}\:
           (( \Gamma( \nu_{x^1\diamond x^2\diamond ... \diamond x^r }(v a )+ 1/2)  \, / \, \Gamma(1/2)) }
{( \Gamma( \bar{\nu}_{x^1\diamond x^2\diamond ... \diamond x^r }(v  )+|A|/2) \,  / \,  \Gamma(|A|/2) )},  \,$$ whereas the definition of $R$  is the same (see (\ref{R}) ).
(Here, as before,   $\bar{\nu}_{x^1\diamond x^2\diamond ... \diamond x^r }(v  )= \sum_{a \in A} \nu_{x^1\diamond x^2\diamond ... \diamond x^r }(v a ). $ Note, that $\bar{\nu}_{x^1\diamond x^2\diamond ... \diamond x^r }(\,)=\sum_{i=1}^r t_i$ if $m=0$.)

The following example is intended to show the difference between
the case  of many samples and one. Let there be two independent
samples $y= y_1 \ldots y_4 = 0101$ and $x = x_1 \ldots x_3 = 101,$
generated by a stationary and ergodic source with the alphabet
$\{0,1\}.$ One wants to estimate the (limiting) probabilities
$P(z_1z_2), z_1,z_2 \in \{0,1\} $ (here $z_1z_2 \ldots $ can be
considered as an independent sequence, generated by the source)
and predict  $x_4x_5$  (i.e. estimate conditional probability
$P(x_4x_5|x_1 \ldots x_3 = 101, y_1 \ldots y_4 = 0101).$ For
solving both problems we will use the measure $R$ (see (\ref{R})).
First we consider the case where $P(z_1z_2)$ is to be estimated
without knowledge of  sequences $x$ and $y.$ From (\ref{Kp}) and
(\ref{km}) we obtain:$$K_0(00)= K_0(11)= \frac{1/2}{1}\,
\frac{3/2}{1+1}= 3/8,\,\: K_0(01)= K_0(10)=\frac{1/2}{1+0}
\,\frac{1/2}{1+1}= 1/8,  $$ $$ K_i(00)= K_i(01)=K_i(10)= K_i(11)=
1/4; \:, \, i \geq 1. $$ Having taken into account the definitions
of $\omega_i$ (\ref{om}) and the measure $R$ (\ref{R}), we can
calculate $R(z_1z_2)$ as follows:
$$ R(00) = \omega_1 K_0(00) + \omega_2 K_1(00) + \ldots = (1- 1/ \log 3)\, 3/8 +(1/ \log 3- 1/ \log 4)\, 1/4 + $$
$$(1/ \log 4- 1/ \log 5)\, 1/4 + \,\ldots = (1- 1/ \log 3)\,\: 3/8 + (1/ \log 3 )\: \:1/4 \approx 0.296 .$$ Analogously, $R(01)= R(10)\approx 0.204, R(11)\approx 0.296. $

Let us now estimate the probability $P(z_1z_2)$ taking into
account that there are two independent samples $y= y_1 \ldots y_4
= 0101$ and $x = x_1 \ldots x_3 = 101.$ First of all we note that
such estimates are based on the formula for conditional
probabilities:
$$ R(z|x\diamond y) = R(x\diamond y \diamond z)/R(x\diamond y) .$$
First we  estimate the frequencies : $$ \nu_{\,0101\diamond 101
}(0)=3, \nu_{\,0101\diamond 101  }(1)=4, \nu_{\,0101\diamond 101
}(00)= \nu_{\,0101\diamond 101  }(11)=0 , \nu_{\,0101\diamond 101
}(01)= 3,$$ $$  \nu_{\,0101\diamond 101  }(10)=2,
\nu_{\,0101\diamond 101  }(010)=1, \nu_{\,0101\diamond 101
}(101)=2, \nu_{\,0101\diamond 101  }(0101)=1, $$ whereas
frequencies of all other tree-letters and four-letters words are
0. Then we calculate :
$$ K_0(\,0101\diamond 101 ) = \frac{1}{2} \frac{3}{4} \frac{5}{6}\frac{7}{8}\frac{1}{10}\frac{3}{12}\frac{5}{14}\approx 0.00244,
K_1(\,0101\diamond 101 ) = (2^{-1})^2 \:\frac{1}{2} \frac{3}{4} \frac{5}{6}\, \:1 \:\frac{1}{2} \frac{3}{4}\, \:1   $$
$$\approx0.0293,\quad K_2(\,0101\diamond 101 ) \approx 0.01172,\quad K_i(\,0101\diamond 101 ) = 2^{-7},\: i\geq 3 ,
$$ $$
R(\,0101\diamond 101 ) = \omega_1 K_0(\,0101\diamond 101 ) + \omega_2 K_1(\,0101\diamond 101 ) + \:\ldots \approx
$$ $$
0.369 \:0.00244 \,+ \, 0.131 \: 0.0293 \,+ 0.06932 \: 0.01172 \:+
2^{-7} \; /\, \log 5  \approx 0.0089. $$ In order to avoid
repetitions, we estimate only one probability $P(z_1z_2= 01).$
Carrying out similar  calculations, we obtain $$ R(0101\diamond
101 \diamond 01) \approx 0.00292,$$ $$ R(z_1z_2= 01|y_1 \ldots y_4
= 0101,x_1 \ldots x_3 = 101) =   R(0101\diamond 101 \diamond 01)/
R(\,0101\diamond 101) \approx 0.32812. $$ If we compare this value
and the estimation $R(01) \approx 0.204$, which is not based on
the knowledge of samples $x$ and $y$, we can see that that the
measure $R$ uses   additional information quite naturally (indeed,
 $01$ is quite frequent in $ y = y_1 \ldots y_4 = 0101$ and $
x= x_1 \ldots x_3 = 101$).

Such generalization can be applied for many universal codes, but,
generally speaking, there exist codes $U$ for which $U(x_1\diamond
x_2)$ is not defined for independent samples $x_1$ and $x_2$ and,
hence, the measure $\mu_U(x_1\diamond x_2)$ is not defined. That
is why we will not describe properties of any universal code, but
for $R$ only. For the measure $R$ all asymptotic properties are
the same for a case of one sample and several ones. More
precisely, the following statement is true:

\textbf{ Claim 2.} \emph{Let $x^1\diamond x^2\diamond ... \diamond
x^r $ be non-overlapping  samples generated by a stationary and
ergodic source and $t$ be a total length of those samples $(t =
\sum_{i=1}^r |x^i| ).$ Then, if $t \rightarrow \infty,$ (and $r$
is fixed) the statements of the Theorems 1-5 are valid, when
applied to $x^1\diamond x^2\diamond ... \diamond x^r $ instead of
the one sample $x_1\ldots x_t.$ (In theorems 2, 4, 5 $\mu_U$
should be changed in $R$.)}

The proofs are analogous to the proofs of the Theorems 1-5.

\section{ Real-valued time series    }

Let $X_t$ be a
 time series with each $X_t$ taking values in  some
interval $\Lambda$. The probability distribution of $X_t$ is
unknown but it is known that the time series  is stationary
and ergodic. 
Let $\{ \Pi_n \}, n \geq 1,$ be an increasing sequence of finite 
partitions that asymptotically generates the Borel sigma-field on
$\Lambda,$ and let $x^{[k]}$ denote the element of $\Pi_k$ that
contains the point $x.$ (Informally, $x^{[k]}$ is obtained by
quantizing $x$ to $k$ bits of precision.)  Suppose that the joint
distribution $P_n$ for $(X_1, \ldots, X_n)$ has a probability
density function $p_n(x_1, \ldots, x_n)$ with respect to a
sigma-finite measure $\lambda_n.$ (For example, $\lambda_n$ can be
Lebesgue measure, counting measure, etc.) For integers $s$ and $n$
we define the following approximation of the density
\begin{equation}\label{ds}
p^s (x_1, \ldots, x_n) =  P(x_1^{[s]}, \ldots, x_n^{[s]}) /
\lambda_n(x_1^{[s]} \ldots x_n^{[s]}).
 \end{equation}

 Let $p(x_{n+1}|x_1, \ldots, x_n)$ denote the conditional density
 given by the ratio $p(x_1, \ldots,x_{n+1}) $ $/ p(x_1, \ldots,x_n)$ for
 $n > 1.$ It is known that for stationary and ergodic processes there exists a
 so-called relative entropy rate $h $ defined  by
\begin{equation}\label{ent}h =   \lim_{n \rightarrow \infty} E(
\log p(x_{n+1}|x_1, \ldots, x_n) ), \end{equation} where $E$ denotes
expectation with respect to $P; $
 see \cite{Ba}.  We also consider
\begin{equation}\label{ents} h_s =   \lim_{n \rightarrow \infty} E(
\log p^s(x_{n+1}|x_1, \ldots, x_n)). \end{equation}

 It is shown by Barron \cite{Ba} that almost surely
 \begin{equation}\label{ba1}  \lim_{t\rightarrow\infty} \frac{1}{t} \log p(x_1 \ldots
 x_t) = h.
\end{equation}
Applying the same  theorem to the density $p^s (x_1, \ldots, x_t),$
we obtain that a.s.
\begin{equation}\label{Ba2}
\lim_{t\rightarrow\infty} \frac{1}{t} \log p^s (x_1, \ldots, x_t) =
h_s.
\end{equation}

Let $U$ be a universal code, which is defined for any finite
alphabet. We define the corresponding density $r_U$ as follows:
\begin{equation}\label{den}    r_U(x_1 \ldots x_t) =
\sum_{i=0}^\infty \omega_i 2^{- |U(x_1^{[i]} \ldots x_t^{[i]})| } /
\lambda_t(x_1^{[i]} \ldots x_t^{[i]})\:.
 \end{equation}
 (It is supposed here that the code $U(x_1^{[i]} \ldots x_t^{[i]})$
 is defined for the alphabet, which contains $|\Pi_i|$ letters.)

It turns out that, in a certain sense,
 the density $r_U(x_1 \ldots x_t)$ estimates the unknown density $p(x_1, \ldots,
 x_t).$

\textbf{Theorem 6 }.  \emph{Let $X_t$ be a stationary ergodic process with
 densities $p(x_1, \ldots , x_t)$ $ = dP_t / d \lambda_t$ such that $
 h  < \infty,$ where  $ h$ is
 relative entropy rate, see  (\ref{ent}). Then
the following equality is true with probability 1:}

\begin{equation}\label{thm}
\lim_{t\rightarrow\infty}\:  \frac{1}{t} \;\{   \log
\frac{p(x_1)}{r_U(x_1)}  \: + \ldots +
  \log \frac{p(x_{n+1}|x_1 ... x_n)}{r_U(x_{n+1}|x_1 ... x_n)}
 \:+   ...+
 \log
\frac{p(x_t|x_1 ... x_{t-1})}{r_U(x_t|x_1 ... x_{t-1})} \: \:
  \}  = 0 . \end{equation}

\textbf{ Proof.} First we note that the following equality can be
easily derived from the definitions and martingale properties:
\begin{equation}\label{limh}
\lim_{s\rightarrow\infty} h_s = h. \end{equation}

 It can be  seen that
(\ref{thm}) is equivalent to the following equality.
\begin{equation}\label{th}
\lim_{t\rightarrow\infty}  \frac{1}{t} \: \log \frac{p(x_1 \ldots
x_t)}{r_U(x_1 \ldots x_t)} \:= \: 0 \: .
 \end{equation}
 First we note that for any integer $s$ the following obvious equality is true:
  $r_U(x_1 \ldots x_t) = \omega_s 2^{- |U(x_1^{[s]} \ldots x_t^{[s]})| } /
\lambda_t(x_1^{[s]} \ldots x_t^{[s]})  \:(1+ \delta)$ for some
$\delta
> 0.$ From this equality and (\ref{th}) we immediately obtain that
a.s.
\begin{equation}\label{th1}
\lim_{t\rightarrow\infty}  \frac{1}{t}  \log \frac{p(x_1 \ldots
x_t)}{r_U(x_1 \ldots x_t)} \, \:\leq  \lim_{t\rightarrow\infty}
\frac{1}{t}
  \log \frac{p(x_1 \ldots x_t)}{2^{-
|U(x_1^{[s]} \ldots x_t^{[s]})| } / \lambda_t(x_1^{[s]} \ldots
x_t^{[s]})  } \, . \end{equation} The right part  can be presented as
follows:
 \begin{equation}\label{th2} \lim_{t\rightarrow\infty} \frac{1}{t} \log \frac{p(x_1 \ldots x_t)}{2^{-
|U(x_1^{[s]} \ldots x_t^{[s]})| } / \lambda_t(x_1^{[s]} \ldots
x_t^{[s]})  } = \lim_{t\rightarrow\infty} \frac{1}{t}
  \log \frac{p^s (x_1, \ldots, x_t)\:\lambda_t(x_1^{[s]} \ldots
x_t^{[s]}) }{2^{- |U(x_1^{[s]} \ldots x_t^{[s]})| } }
\end{equation}
$$ + \lim_{t\rightarrow\infty} \frac{1}{t} \log \frac{p(x_1 \ldots x_t)}{p^s (x_1, \ldots,
x_t)}.
$$
Having taken into account  that $U$ is the universal code and
(\ref{ds}), we can see that the  first term  equals to zero. From
(\ref{ba1})
 and (\ref{Ba2})  we can see that a.s. the second term is
equal to $h - h_s.$ This equality is valid for any integer $s$
and, according to (\ref{limh}), $\lim_{s\rightarrow\infty} h_s =
h.$ Hence, the second term equals to zero, too, and we obtain the
proof of  (\ref{th}). The  theorem is proven.

\textbf{Corollary 1. }
$$
  \lim_{t\rightarrow\infty}  \frac{1}{t} E( \log  \frac{p(x_1 ... x_t)}{r_U(x_1 ... x_t)} )    = 0.
$$
\textbf{ Proof.}  Analogously to      (\ref{th1})  and (\ref{th2}) we can obtain the following enequality
\begin{equation}\label{th5}
   \log \frac{p(x_1 \ldots
x_t)}{r_U(x_1 \ldots x_t)} \, \:\leq
 \log \frac{p(x_1 \ldots x_t)}{2^{-
|U(x_1^{[s]} \ldots x_t^{[s]})| } / \lambda_t(x_1^{[s]} \ldots
x_t^{[s]})  } =
  \log \frac{p^s_t (x_1, \ldots, x_t)\:\lambda_t(x_1^{[s]} \ldots
x_t^{[s]}) }{2^{- |U(x_1^{[s]} \ldots x_t^{[s]})| } }
\end{equation}
$$ +  \log \frac{p(x_1 \ldots x_t)}{p^s (x_1, \ldots,
x_t)}.
$$ for any integer $s.$
Hence,
\begin{equation}\label{th6}
 \frac{1}{t} E (  \log \frac{p(x_1 \ldots
x_t)}{r_U(x_1 \ldots x_t)} \,) \:\leq E(
\frac{1}{t}
  \log \frac{p^s_t (x_1, \ldots, x_t)\:\lambda_t(x_1^{[s]} \ldots
x_t^{[s]}) }{2^{- |U(x_1^{[s]} \ldots x_t^{[s]})| } } )  + E(
\frac{1}{t} \log \frac{p(x_1 \ldots x_t)}{p^s (x_1, \ldots, x_t)})
.  \end{equation} The first term  is the average redundancy of a
universal code for a finite-alphabet source,  hence, it  tends to
0 according to the definition of the universal code. The second
term tends to      $h - h_s$ for any $s,$ hence, it is equals to
zero. Corollary 1 is proven.

\textbf{Corollary 2. }
$$i)
\lim_{t\rightarrow\infty}  \frac{1}{t} \int  (\, p(x_1 \ldots x_t)
- r_U(x_1 \ldots x_t)\, )^2 \,d \lambda_t \:= \: 0 ,
$$ $$
ii) \lim_{t\rightarrow\infty}  \frac{1}{t} \int |\, p(x_1 \ldots
x_t) - r_U(x_1 \ldots x_t)\,| \,d \lambda_t \:= \: 0 .
$$

\textbf{ Proof}  i)     immediately follows from the corollary 1
and the Pinsker's inequality   (\ref{pi}). ii) can be derived from
i) and the Jensen inequality for the function $x^2.$

\textbf{Theorem 7 }.     \emph{Let $B_1, B_2, ... $ be a  sequence
of  measurable    sets. Then the following equalities  are true:}
\begin{equation}\label{thc} i)\:
\lim_{t\rightarrow\infty}   \: E( \frac{1}{t} \sum_{m=0}^{t-1} (
P(x_{m+1} \in B_{m+1}  | x_1 ... x_m) -   R_U(x_{m+1} \in B_{m+1}
| x_1 ... x_m) )^2 ) = 0\,,
\end{equation}
$$ ii) \:
E( \frac{1}{t} \sum_{m=0}^{t-1} | P(x_{m+1} \in B_{m+1}  | x_1 ...
x_m) -   R_U(x_{m+1} \in B_{m+1}  | x_1 ... x_m) )| = 0\,.
$$
\textbf{ Proof.}  Obviously,
 \begin{equation}\label{tha}
 E( \frac{1}{t} \sum_{m=0}^{t-1} ( P(x_{m+1} \in B_{m+1}  | x_1 ... x_m) -   R_U(x_{m+1} \in B_{m+1}  | x_1 ... x_m) )^2 ) \leq
\end{equation}
$$ \frac{1}{t} \sum_{m=0}^{t-1} E ( | P(x_{m+1} \in B_{m+1}  | x_1 ... x_m) -   R_U(x_{m+1} \in B_{m+1}  | x_1 ... x_m) |
+ $$
$$  | P(x_{m+1} \in \bar{ B}_{m+1}  | x_1 ... x_m) -   R_U(x_{m+1} \in\bar{ B}_{m+1}  | x_1 ... x_m) | )^2 . $$
From the Pinsker inequality     (\ref{pi}) and    convexity of the
KL divergence   (\ref{r0}) we obtain the following inequalities
\begin{equation}\label{thb}
 \frac{1}{t} \sum_{m=0}^{t-1} E ( | P(x_{m+1} \in B_{m+1}  | x_1 ... x_m) -   R_U(x_{m+1} \in B_{m+1}  | x_1 ... x_m) |
+
\end{equation}
$$  | P(x_{m+1} \in \bar{ B}_{m+1}  | x_1 ... x_m) -   R_U(x_{m+1} \in\bar{ B}_{m+1}  | x_1 ... x_m) | )^2 \leq $$
$$ \frac{ const}{t} \sum_{m=0}^{t-1} E( ( \log \frac{ P(x_{m+1} \in B_{m+1}  | x_1 ... x_m)}{ R_U(x_{m+1} \in B_{m+1}  | x_1 ... x_m)} +
 \log  \frac{  P(x_{m+1} \in \bar{ B}_{m+1}  | x_1 ... x_m) }  { R_U(x_{m+1} \in\bar{ B}_{m+1}  | x_1 ... x_m) } )  \leq $$
$$  \frac{const}{t} \sum_{m=0}^{t-1} ( \int   p( x_1 ... x_m) (\int p(  x_{m+1} | x_1 ... x_m) ) \log  \frac{p(  x_{m+1} | x_1 ... x_m)
} {r_U(  x_{m+1} | x_1 ... x_m) } d \lambda ) d \lambda_m ) .$$
Having taken into account that     the last term is equal to $
\frac{const}{t}  E( \log \frac{ p(x_1 ... x_t) } { r_U( x_1 ...
x_t) } ) , $  from     (\ref{tha})  and (\ref{thb}) and  Corollary
1 we obtain   (\ref{thc}). ii) can be derived from i) and the
Jensen inequality for the function $x^2.$ The theorem is proven.

We have seen that in a certain sense the estimation  $r_U$
approximates the density $p.$  The following theorem shows that  $r_U$
can be used instead of $p$ for estimation of average values of certain functions.

\textbf{Theorem 8 }.     \emph{Let $f$ be   an integrable
function, whose absolute value is bounded  by a certain constant
$M$. Then the following equalities are valid:}
\begin{equation}\label{int}
i) \lim_{t\rightarrow\infty}   \frac{1}{t} E( \sum_{m=0}^{t-1}  (
\int  f(x) p(x  | x_1 ... x_m) d \lambda_m  -  \int  f(x) r_U(x  |
x_1 ... x_m) d \lambda_m  )^2 )  = 0,
\end{equation}
$$ ii) \:
\lim_{t\rightarrow\infty}   \frac{1}{t} E( \sum_{m=0}^{t-1}  |
\int  f(x) p(x  | x_1 ... x_m) d \lambda_m  -  \int  f(x) r_U(x  |
x_1 ... x_m) d \lambda_m  | )  = 0.$$

\textbf{ Proof.}  The last inequality from the following chain
follows from the Pinsker's one, whereas all others are obvious.
$$    ( \int  f(x) p(x  | x_1 ... x_m) d \lambda_m  -  \int  f(x) r_U(x  | x_1 ... x_m)
d \lambda_m  )^2   = $$ $$   ( \int  f(x) (p(x  | x_1 ... x_m)  - r_U(x  | x_1 ... x_m) ) d \lambda_m  )^2     \leq
M^2  ( \int  (p(x  | x_1 ... x_m)  - r_U(x  | x_1 ... x_m) ) d \lambda_m  )^2
$$   $$
\leq    M^2  ( \int  |p(x  | x_1 ... x_m)  - r_U(x  | x_1 ... x_m)
|  d \lambda_m  )^2 \leq $$ $$ const  \,   \int    p(x  | x_1 ...
x_m) \log (p(x  | x_1 ... x_m)/r_U(x  | x_1 ... x_m) d \lambda_m
.
$$
From these inequalities we obtain:
\begin{equation}\label{int1}
 \sum_{m=0}^{t-1}   E (  \int  f(x) p(x  | x_1 ... x_m) d \lambda_m  -  \int  f(x) r_U(x  | x_1 ... x_m)
d \lambda_m  )^2  )  \leq
\end{equation}
 $$  \sum_{m=0}^{t-1}
const  \,    E ( \int    p(x  | x_1 ... x_m) \log (p(x  | x_1 ...
x_m)/r_U(x  | x_1 ... x_m) ) d \lambda_m   .
$$
The last term can be presented as follows:
$$
\sum_{m=0}^{t-1}   E ( \int    p(x  | x_1 ... x_m) \log (p(x  | x_1 ... x_m)/r_U(x  | x_1 ... x_m) ) d \lambda_m  )  = $$
$$            \sum_{m=0}^{t-1}   \int     p(x_1 ... x_m)
 \int    p(x  | x_1 ... x_m) \log (p(x  | x_1 ... x_m)/r_U(x  | x_1 ... x_m))  d \lambda \, d \lambda_m ) =  $$
$$  \int    p(x_1 ... x_t) \log (p(x_1 ... x_t)/r_U(x_1 ... x_t))  d \lambda_t     . $$
From this equality,    (\ref{int1}) and   Corollary 1 we obtain
(\ref{int}). ii) can be derived from (\ref{int1}) and the Jensen
inequality for  $x^2.$ Theorem is proven.

\section{The Experiments }

In this part we describe the results of some experiments and a
simulation study carried out in order to evaluate the
 efficiency of the suggested algorithms, paying the
main attention to the prediction problem. The obtained results
show that, in general, the described
 approach  can be used in
applications.

\subsection{ Simulations}

We constructed several artificial samples created by processes
with known structure and tried to predict the next value
($x_{n+1}$)  of the process based on $x_1, ..., x_n$.
 WinRAR archiver (http://www.rarlab.com)
was chosen as a code for constructing predictors. The scheme of
experiments is as follows. Let $x_{1}\dots x_{n}$ be the generated
sequence.  Denote by $x^*$ the estimation of $x_{n+1}$. For each
$n$ we calculate the density $r_U(x_{1}\dots x_{n})$ and the
average value (according to this density), which is output as the
predicted value $x^*$.

The first process was created according to  the following formula:
$ x_{i} = \sin\left(\pi*i/23\right) .$
 In this experiment we used
WinRAR with the medium quality of
compression.\\
After every experiment the error of the prediction
$r_{i}=|x^*-x_{n+1}|$ was evaluated. We compared these values with
errors of the so-called  inertial predictor, where the estimation
of (unknown) $x_{n+1}$ is defined as $x_{n}$ (i.e. $x^* = x_{n}$).
The obtained results are given in the following table.

\begin{table}[h]


\begin{tabular}{|c|c|c|c|}
  \hline
  Number of experiments&Length of a sample sequence (n)&Suggested&Inertial\\
  \hline
  100 & 1000 & 0.37 & 0.41 \\
  \hline
  100 & 2000 & 0.37 & 0.46 \\
  \hline
  100 & 3000 & 0.34 & 0.45 \\
  \hline
\end{tabular}

\end{table}

 The numbers given in the first line of the table mean that 100
experiments were carried out, the length of the observed  data is
equal to 1000 ($n=1000$), the mean value of the error $ (
\sum_{i=1}^{100}r_{i}/100 ) $ of prediction using the suggested
method is  0.37, whereas  the mean value of inertial prediction is
0.41.

The second  was a "random mixture" of the   four following
functions:
$f_{1}(i)= \left[5*\sin\left(\pi*i/16\right)\right]$,
$f_{2}(i)= \left[7*\sin\left(\pi*i/+\pi/5\right)\right]$,
$f_{3}(i)= \left[8*\sin\left(\pi*i/3\right)\right]$,
$f_{4}(i)= \left[8*\sin\left(\pi*i/23\right)\right]$. More precisely, first the length of a segment was randomly chosen according to the Poisson distribution (with a parameter $\lambda = 0.1$), then the  function on each
segment was chosen randomly  (with the probability $1/4$) and values of the segment were generated according to the chosen  formula.
The results of this experiment are given in the table below. \\ \\
\begin{tabular}{|c|c|c|c|}
  \hline
  Number of experiments&Length of a sample sequence (n)&Suggested&Inertial\\
  \hline
  100 & 2000 & 1.43 & 2.2 \\
  \hline
  100 & 5000 & 2.97 & 4.27 \\
  \hline
  100 & 10000 & 3.07 & 3.4 \\
  \hline
\end{tabular}\\ \\

\subsection{ Prediction of currency rate}
To carry out  this experiments we took values EURO/USD from Forex
stock \\ (http://www.forex.com). The scheme of the experiments is
mainly the same as in the previous section. In these experiments
we used WinRAR data compression method and predictor $R$. First of
all we carried out few experiments to find best parameters for
prediction.  $R$ showed better results than WinRAR archiver. We
took independent samples and carried out experiments as it was
described above.
The results are given in the table below.\\
\\
\begin{tabular}{|c|c|c|c|c|c|}
\hline
Number of experiments&Length of a sample sequence (n)&Suggested&Inertial\\
\hline
100 & 600 & 0.0150 & 0.0175\\
\hline
100 & 600 & 0.0143 & 0.0165\\
\hline
100 & 600 & 0.0131 & 0.0162\\
\hline
100 & 600 & 0.0164 & 0.0175\\
\hline
\end{tabular}\\ \\

So, we can see that predictors which are based on data compression
methods have reasonable performance in practice.

\end{document}